\documentclass[aps,
    pra,
    twocolumn, 
    longbibliography=false]{revtex4-2}
\usepackage{siunitx}
\usepackage{amsmath,amssymb,mathrsfs}
\usepackage{psfrag}
\usepackage{graphicx}
\usepackage{graphics}
\usepackage{epsfig}
\usepackage{bm}
\usepackage{color}
\usepackage{verbatim,color}
\usepackage{physics}
\usepackage[normalem]{ulem}
\usepackage{cancel}
\usepackage[colorlinks=true,linkcolor=blue,citecolor=black,urlcolor=blue]{hyperref}

\usepackage[acronym]{glossaries}
\newacronym{dr}{DR}{dimensional regularization}
\newacronym{ms}{MS}{minimal subtraction}
\newacronym{eft}{EFT}{effective field theory}
\newacronym{mqst}{MQST}{macroscopic quantum self-trapping }
\newacronym{lhy}{LHY}{Lee-Huang-Yang}
\newcommand{\beq}{\begin{equation}}
\newcommand{\eeq}{\end{equation}}
\newcommand{\beqa}{\begin{eqnarray}}
\newcommand{\eeqa}{\end{eqnarray}}
\newcommand{\ba}{\begin{aligned}[b]}
\newcommand{\ea}{\end{aligned}}

\definecolor{darkgreen}{rgb}{0.0, 0.5, 0.0}
\definecolor{shadow}{rgb}{0.6, 0.6, 0.6}
\definecolor{darkgreen}{rgb}{0.0, 0.5, 0.0}
\definecolor{c1}{rgb}{0.9, 0.0, 0.0}
\definecolor{c2}{rgb}{0.3, 0.0, 0.6}
\definecolor{c3}{rgb}{0.9, 0.0, 0.0}

\newcommand{\mn}[1]{\marginpar{\tiny}}
\usepackage{tikz}

\definecolor{plotgold}{rgb}{0.85, 0.65, 0.13}
\definecolor{plotorange}{rgb}{0.9, 0.45, 0.0}
\definecolor{plotpurple}{rgb}{0.5, 0.0, 0.5}
\definecolor{plotalbiez}{rgb}{0.9, 0.45, 0.0}
\definecolor{plotgreen}{rgb}{0.0, 0.45, 0.0}

\newcommand{\plotmarkcircle}[1]{\tikz[baseline=-0.6ex]\filldraw[fill=#1,draw=black,line width=0.3pt] (0,0) circle (2.1pt);}
\newcommand{\plotmarksquare}[1]{\tikz[baseline=-0.6ex]\filldraw[fill=#1,draw=black,line width=0.3pt] (-2.1pt,-2.1pt) rectangle (2.1pt,2.1pt);}
\newcommand{\plotmarkdiamond}[1]{\tikz[baseline=-0.6ex]\filldraw[fill=#1,draw=black,line width=0.3pt] (0,2.6pt) -- (2.6pt,0) -- (0,-2.6pt) -- (-2.6pt,0) -- cycle;}
\newcommand{\plotmarktriangle}[1]{\tikz[baseline=-0.6ex]\filldraw[fill=#1,draw=black,line width=0.3pt] (-2.6pt,-2.0pt) -- (2.6pt,-2.0pt) -- (0,2.6pt) -- cycle;}

\newcommand{\markSingh}{\plotmarkcircle{blue}}
\newcommand{\markPezze}{\plotmarksquare{plotgreen}}
\newcommand{\markBernhart}{\plotmarkdiamond{plotpurple}}
\newcommand{\markAlbiez}{\plotmarktriangle{plotalbiez}}

\begin{document}
\title{Bosonic Josephson junction dynamics:\\ interplay between quantum and thermal fluctuations}
\author{A. Bardin$^{1,3}$, F. Lorenzi$^{2}$, and L. Salasnich$^{1,3,4}$}
\affiliation{$^{1}$Dipartimento di Fisica e Astronomia "Galileo Galilei", 
Universit\`a di Padova, Via Marzolo 8, 35131 Padova, Italy\\
$^{2}$ Dipartimento di Ingegneria dell'Informazione, Università di Padova, Via Gradenigo 6A, 35131 Padova, Italy \\
$^{3}$Istituto Nazionale di Fisica Nucleare (INFN), Sezione di Padova, via Marzolo 8, 35131 Padova, Italy\\
$^{4}$Istituto Nazionale di Ottica (INO) del Consiglio Nazionale delle Ricerche (CNR), via Nello Carrara 1, 50019 Sesto Fiorentino, Italy}

\date{\today}
\begin{abstract}
We investigate the superfluid dynamics of a Josephson junction beyond the mean-field description, incorporating the role of thermal fluctuations as well as quantum fluctuations. Using a formalism that accounts for the fluctuations in a homogeneous gas, and under the assumption that the transport of the non-condensed component is negligible, we derive a corrected equation of motion within the two-site approximation. The resulting corrections for the typical dynamical quantities, like the Josephson frequency, the strength of macroscopic quantum self-trapping, and the threshold for spontaneous symmetry breaking, allow us to predict the effects of both types of fluctuations and assess their relative importance in different regimes  in a semianalytical fashion. For all the dynamical quantities, the quantum fluctuations are shown to play an opposite role with respect to the thermal fluctuations. Josephson frequency is decreased by thermal fluctuations and both the critical strenghts of macroscopic quantum self trapping and spontaneous symmetry breaking are increased. We assess the experimentally accessible regimes by calculating the relevant parameters of recent experimental realizations of Bosonic Josephson junction and show that the expected regime is dominated by quantum fluctuations.
\end{abstract}
\maketitle

\section{Introduction} 
Josephson junctions are macroscopic systems that exhibit purely quantum phenomena. 
They consist of two superconductors, or superfluids that are weakly coupled \cite{josephson_possible_1962}. The junction can be realized with superconductors separated by a thin insulator \cite{anderson_probable_1963}, and, as has been shown more recently, also with Bose-condensed gases of alkali atoms either separated by a thin potential created by a ``sheet" of laser light \cite{albiez_direct_2005, levy_c_2007, shin_atom_2004}, or loaded into an optical lattice as an array of Josephson junctions \cite{anderson_macroscopic_1998, cataliotti_josephson_2001}. 

The atomic Josephson junction is not limited to bosonic species. It is indeed possible to study the case of fermionic superfluids across the BEC-BCS crossover, with which Josephson junctions have been realized \cite{valtolina_josephson_2015, del_pace_tunneling_2021, kwon_strongly_2020, burchianti_connecting_2018} and explained theoretically \cite{spuntarelli_josephson_2007,zaccantizwerger,wlazlowski_dissipation_2023, piselli_josephson_2020}.
Atomic Josephson junctions are gaining interest as a technological platform for systems using  optical lattices, and they have been proposed as a platform for quantum computing \cite{tian_quantum_2003} and atomtronic devices \cite{amico_colloquium_2022, amico_roadmap_2021}.
Limiting ourselves to the bosonic case, the possibility to set an almost arbitrary external potential has led to the realization of many different spatial configurations of these systems. The configurations where the size of each region crucially affects the dynamics are commonly referred to as extended bosonic Josephson junctions (BJJ).
An effective theoretical approach to the analysis of the BJJ is the two-mode approximation, where the spatial structure of the superfluid is discarded, obtaining an effective zero-dimensional model with two coupled variables \cite{raghavan_coherent_1999, smerzi_quantum_1997}. These are the canonically conjugated pair of population imbalance and relative phase operators \cite{leggett_concept_1991}. %
While this approach has drawbacks, being limited to sufficiently weak interactions \cite{burchianti_josephson_2017, xhani_dynamical_2020}, it still allows for the effective description of a lot of dynamical features of the junctions, both in mean-field \cite{smerzi_quantum_1997} and beyond mean field \cite{furutani_quantum_2021, furutani_interaction-induced_2024, wimberger_finite-size_2021}.
Within this approach, it is possible to naturally describe two distinct dynamical regimes \cite{smerzi_quantum_1997, raghavan_coherent_1999}. The first is the Josephson regime, characterized by sinusoidal oscillations of both atom quantities around a zero mean value in the deep tunneling regime. The oscillation frequency in this case is known as the Josephson frequency. 
The second regime, known as macroscopic quantum self-trapping (MQST), features oscillations of the population imbalance with a small amplitude around a nonzero mean value, while the relative phase increases over time at an average rate known as the phase-slippage rate \cite{avenel_josephson_1988}.
Another known  phenomenon is found by considering the phase difference between the sites fixed at $\pi$. In this case, when the on-site interaction is greater than a given value, two more fixed points are found at population imbalances  $\pm  z_s$, in addition to the one at $z=0$. This constitutes a spontaneous symmetry breaking (SSB), which can be understood as a bifurcation in the energy landscape of the two-mode phase space \cite{raghavan_coherent_1999}.

The inclusion of the effects of temperature brings novel difficulties related to the treatment of the non-condensed fraction and the complex many body dynamics of the condensed fraction. A full numerical account of the mean-field Gross--Pitaevskii equation (GPE) dynamics of the condensed fraction coupled with classical hydrodynamical equations of the noncondensed fraction can be performed within the Zaremba-Nikuni-Griffin (ZNG) formalism \cite{zng, xhani_dissipation_2022, xhani_dynamical_2020}. While powerful, this approach rarely yields analytical estimates. In this Letter, we extend the two-mode formalism by formulating the quantum and thermal corrections to the well-known dynamical features of the Josephson junction discussed previously for a three-dimensional Josephson junction.
In contrast to the ZNG approach, the use of homogeneous quantum fluctuations in both sites leads to effective two-mode dynamics that can be used to obtain analytical estimates for the dynamical features of the junction.
Our method, unlike previous methods addressing the finite temperature dynamics using the stochastic projected GPE \cite{bidasyuk_finite-temperature_2018}, or with the two-site von Neumann equation for the density matrix in absence of interactions  \cite{korshynska_generalized_2024}, is based  on the approximation of the thermal and quantum fluctuations of the two superfluids with those of a homgeneous gas \cite{salasnich_zero-point_2016}, in a similar way as discussed in Ref.~\cite{bardin_quantum_2024} for the sole quantum fluctuations.

The quantities we focus on are the Josephson frequency, the macroscopic quantum self trapping critical strength, and the strength of the onset of spontaneous symmetry breaking. We obtain expressions of their dependence on the temperature and the gas parameter.
Each of the dynamical quantities shows a competition effect between thermal and quantum fluctuations. In particular, the effect of temperature is shown to decrease the Josephson frequency and to increase the MQST and SSB critical strengths. The quantum fluctuations play the opposite role in all the dynamical quantities. This shows that, within the limits of our approximation, a regime where the effects cancel each other is present.

The article is divided into the following sections: in Section II, we define the relevant physical quantities and introduce the modified two-mode formalism. In Section III, we obtain the corrected Josephson frequency; in Section IV, we describe the modifications to the SSB, and in Section V, the one to the MQST. We discuss the relevance of our predictions, and we conclude in Section VI.

In the Appendices, we review the thermodynamics of a uniform ultracold Bose gas, and we outline two asymptotic regimes for low and high temperatures, leading to closed-form estimates of the dynamical quantities. They are split in Appendix A for the calculations related to the condensate fraction and Appendix B for the ones related to the chemical potential.

\section{Two-site model for the Josephson dynamics}
\begin{figure}[!htbp]
    \centering
    \includegraphics[width=0.8\linewidth]{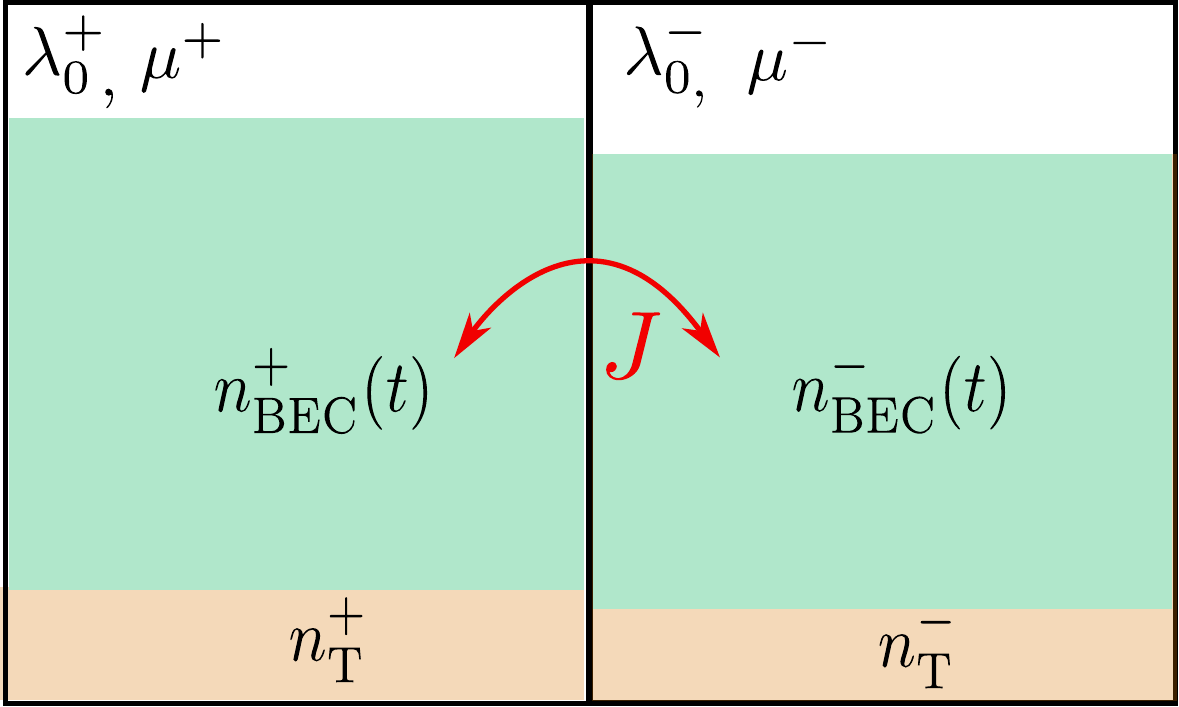}
    \caption{
    Schematic of a bosonic Josephson junction at finite temperature.
    Two regions (labeled $+$ and $-$) contain both a condensed fraction with time-dependent density $n_{\mathrm{BEC}}^{\pm}(t)$ and a thermal component with density $n_T^{\pm}$.
    Each subsystem is characterized by its chemical potential $\mu^{\pm}$ and the condensate fraction $\lambda_0^{\pm}$.
    Tunneling of condensed atoms across the barrier is described by the coupling $J$.
    }
    \label{fig:scheme}
\end{figure}
Let us define a two-sites, three-dimensional model of a bosonic Josephson junction consisting of two sites, labeled as ``$\pm$", of equal volume $L^3$ at constant temperature $T$; each of these sites contains a repulsive bosonic gas of ultracold and dilute neutral atoms of density $n^{\pm}\equiv N^\pm/L^3$ and chemical potential $\mu^{\pm}\equiv\mu(n^\pm,g,T)$ (see Appendix \ref{Appendix B} for more details) provided by a contact-like interaction, characterized by an interaction coefficient $g=4\pi\hbar^2 a_s/m$, where $a_s$ is the $s$-wave scattering length and $m$ is the particle mass of the gas, on each site. Particles can tunnel from one site to the other through coherent tunneling of the condensate fraction \cite{zaccantizwerger,PhysRevA.102.013325}, mediated by the coupling constant $J$. The  total number of particles is conserved, and so is the total number density $n=(n^+(t)+n^-(t))/2$, even if the densities on each site vary with time. In the following, we assume that the total number of particles in the condensate, and therefore the total condensate number density $n_{BEC}=(n^+_{BEC}(t)+n^-_{BEC}(t))/2$, is constant. This assumption, along with the conservation of the total number of particles, implies that the total number of non-condensed atoms is constant. Moreover, we assume that the numbers of non-condensed atoms in the two sites are stationary, namely that  the thermal particle density $n^\pm_{th}$ is constant in time.   A schematic representation of our system is shown in Fig.~\ref{fig:scheme}. 
This assumption is reasonable for experiments working with large condensed fraction, at least in a short time span from the beginning of the dynamics.
 %
 A consequence of this assumption is that the total condensate density $n_{BEC}$, or equivalently the total condensate fraction $\lambda_T=n_{BEC}/n$, is fixed once the initial total population imbalance $z_{tot,0}\equiv z_{tot}(0)$ and the initial condensate fraction $\lambda_0^\pm=n_{BEC}^\pm/n^\pm\equiv \lambda(n^\pm,g,T)$ are known. In particular, the total condensate fraction at the time $t=0$ can be computed from the condensate fractions on the two sites, taking into account the imbalance is given by
 \begin{equation}
    \lambda_T=\frac{\lambda_0^++\lambda_0^-}{2}+z_{tot,0}\frac{\lambda_0^+-\lambda_0^-}{2}
\end{equation}
Moreover, the initial condensate population imbalance $z_{0}\equiv z(0)$ is given by the relation
\begin{equation}
    \lambda_Tz_{0}=\frac{\lambda_0^+-\lambda_0^-}{2}+z_{tot,0}\frac{\lambda_0^++\lambda_0^-}{2}
\end{equation}
 We can now present a generalized version of the Josephson-Smerzi \cite{raghavan_coherent_1999, smerzi_quantum_1997} equations of motion which accounts for the quantum and thermal fluctuations.
The equations of motion  for the condensate population imbalance $z=(n_{BEC}^+-n_{BEC}^-)/2n_{BEC}$ and the condensate relative phase $\phi = \phi_--\phi_+$ are given by
\begin{equation}\label{eom-smerzi}
    \begin{split}
        \dot z&=-\frac{J}{\hbar}\sqrt{1-z^2}\sin\phi\\
        \dot \phi&=\frac{J}{\hbar}\frac{z}{\sqrt{1-z^2}}\cos\phi+\frac{\mu_+-\mu_-}{2\hbar}
    \end{split}
\end{equation}
which differs from the usual Josephson-Smerzi equations because instead of the term $2gnz$ (see Eq.  11 of Ref.~\cite{bardin_quantum_2024}), clearly linear in the population imbalance $z$, we include the difference of the beyond-mean-field chemical potentials between the two sites, i.e. $\mu_+-\mu_-$, in such a way that quantum and thermal fluctuations are taken into account. Another change compared to the mean-field equations is the tunneling coupling constant, as pointed out by \cite{zaccantizwerger,PhysRevA.102.013325}, both the condensate fraction and the chemical potential at equilibrium play a role in the tunneling rate between the two sites; therefore, we consider $J\equiv J_0\lambda_T\sqrt{\bar\mu_{eq}}$, where $J_0$ refers to the mean-field tunneling coupling, $\lambda_T\equiv n_{BEC}/n$ is the system condensate fraction and $\bar\mu_{eq}=\mu_{eq}/gn$ is the beyond-mean-field chemical potential at equilibrium, i.e. $z_{tot,0}=0$, renormalized on the mean-field one.
Defining $\bar\mu_\pm\equiv\mu_\pm /gn$ we can rewrite the equations of motion \eqref{eom-smerzi} as
\begin{equation}
    \begin{split}
        \dot z&=-\frac{J_0\lambda_T\sqrt{\bar\mu_{eq}}}{\hbar}\sqrt{1-z^2}\sin\phi\\
        \dot \phi&=\frac{J_0\lambda_T\sqrt{\bar\mu_{eq}}}{\hbar}\left(\frac{z}{\sqrt{1-z^2}}\cos\phi+\frac{\Lambda}{\lambda_T\sqrt{\bar\mu_{eq}}}\frac{\bar\mu_+-\bar\mu_-}{2}\right)
    \end{split}\label{eq:smerzi-nostre}
\end{equation}
where $\Lambda=gn/J_0$ is the critical strength. 
Note that, in our system, there is the following subtlety: only a fraction of the particles can tunnel, while both the condensate and the non-condensate parts contribute to the chemical potential. Namely, we have that the chemical potentials of the two sites $\mu_\pm=\mu(n^\pm,g,T)$ are functions of the total number of particles in each site and, with a bit of elementary algebra, we can express the two chemical potentials as a function of the condensate population imbalance $z(t)$ such as
\begin{equation}
    \mu_\pm=\mu(1\pm\lambda_T(z(t)-z_0)\pm z_{tot,0}) \,.
\end{equation}
We find that the initial population imbalances, together with the condensate fraction, set an offset to the arguments of the chemical potentials. 

\section{Josephson regime}
\begin{figure}[!htbp]
    \centering
    \includegraphics[width=\linewidth]{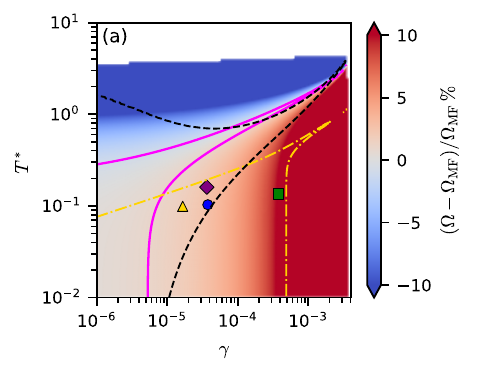}
    \includegraphics[width=\linewidth]{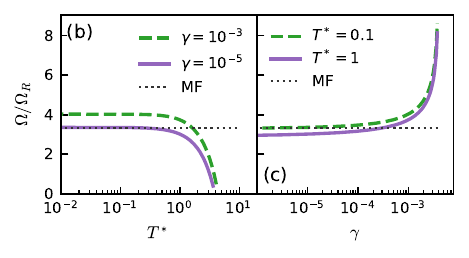}
    \includegraphics[width=\linewidth]{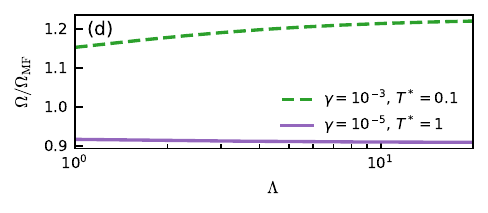}
    \caption{Josephson frequency corrections. (a) Colormap of the relative beyond-mean-field correction of the Josephson frequency \ensuremath{\Omega} for a bosonic gas at finite temperature \ensuremath{T^* =mk_BT/\hbar^2n^{2/3}}, gas parameter \ensuremath{\gamma=a_s^3n} and strength parameter \ensuremath{\Lambda=10}. The bold magenta line marks \ensuremath{1\%} relative correction; the black dashed and yellow dashed-dotted lines indicate the breakdown of the high- and low-temperature approximations, respectively. The four experimental points correspond to \cite{albiez_direct_2005} (orange triangle), \cite{pezze2024} (green square), \cite{singh2025} (blue circle) and \cite{bernhart2024} (purple diamond), with values listed in Table \ref{tab:comparison}. Panel (b) is \ensuremath{\Omega/\Omega_{\mathrm{R}}}, with $\Omega_R=J_0/\hbar$, as a function of  \ensuremath{T^*} at fixed \ensuremath{\gamma}, and panel (c) as a function of  \ensuremath{\gamma} at fixed \ensuremath{T^*}. Panel (d) shows the dependence of the correction on $\Lambda$ for fixed choices of $\gamma$ and $T^*$.}
    \label{fig:Omega}
\end{figure}

The Josephson regime is the dynamical regime in which, at the mean-field level, the condensate population imbalance $z$ and the phase difference $\phi$ follow the equations of motion of a harmonic oscillator; the characteristic frequency is then called the Josephson frequency, and at the mean-field level it is given by $\Omega_{\text{MF}}=\Omega_R\sqrt{1+\Lambda}$ with $\Omega_R=J_0/\hbar$ is the Rabi frequency. The system is in the Josephson regime if the initial conditions, namely the initial population imbalance $z_0$ and the initial phase difference $\phi_0$ are close to the equilibrium ones, i.e. $(z_{eq},\phi_{eq})=(0,0)$. Considering now the limit of low total population imbalance, i.e. $z_{tot,0}\to 0$, we can approximate the system condensate fraction $\lambda_T$ as constant, i.e.
\begin{equation}
    \lambda_T=\lambda_{eq}+o(z^2_{tot})
\end{equation}
where $\lambda_{eq}=\lambda(n,g,T)$ and the initial condensate population imbalance $z_{0}$ as linear in $z_{tot,0}$, i.e.
\begin{equation}
\lambda_Tz_{0}=z_{tot,0}\left(\lambda_{eq}+\lambda'_{eq}  \right)+o(z^3_{tot})
\end{equation}
where $\lambda'_{eq}\equiv \partial\lambda(n(1+z_{tot, 0}),g,T)/\partial z_{tot,0}\Big|_{z_{tot,0}=0}$. Linearizing the chemical potential difference around $z(t)\rightarrow0$ and then expanding in powers of $z_{tot,0}$, we find the following relation
\begin{equation}
    \frac{\mu_+-\mu_-}{2}=\lambda_{eq}\mu'_{eq}\left[z(t)-z_{tot,0}\frac{\lambda_{eq}+\lambda'_{eq}-1}{\lambda_{eq}}\right] \,.
\end{equation}
In a similar way to $\lambda'_{eq}$, $\mu'_{eq}$ is defined as $\mu'_{eq}\equiv \partial\mu_{eq}(n(1+z_{tot, 0}),g,T)/\partial z_{tot,0}\Big|_{z_{tot,0}=0}$, and $\bar{\mu}'_{eq} = \mu'_{eq}/(gn)$.
The solutions of the linearized equations of motion are then harmonic, and in the form
\begin{equation}
    \begin{split}
        z(t)&=(z_0-\bar z)\cos{\Omega t}+\bar z\\
        \phi(t)&=\phi_0\cos{\Omega t}
    \end{split}
\end{equation}
with frequency $\Omega$ given by
\begin{equation}
    \Omega=\frac{J_0}{\hbar}\lambda_{eq}\bar\mu^{1/2}_{eq}\sqrt{1+\frac{\bar\mu'_{eq}}{\bar\mu^{1/2}_{eq}}\Lambda}
\end{equation}
and equilibrium shift 
\begin{equation}
    \bar z=z_{tot,0}\frac{\lambda_{eq}+\lambda'_{eq}-1}{\lambda_{eq}}\frac{\Lambda\bar\mu'_{eq}}{\Lambda\bar\mu'_{eq}+\bar\mu^{1/2}_{eq}}
\end{equation}
which is linear in $z_{tot,0}$ and vanishes for $\Lambda\rightarrow0$, while for large $\Lambda$ it approaches $\bar z=(z_0-z_{tot,0})/\lambda_{eq}$. We observe that, while the phase difference \(\phi\) undergoes harmonic oscillations with frequency \(\Omega\) around zero as in the mean-field case, the population imbalance \(z(t)\) instead oscillates around a shifted center \(\bar{z}\) with amplitude \(z_0 - \bar{z}\). 
In Fig.~\ref{fig:Omega}, we show the calculations of the Josephson frequency. These calculations are performed using the numerically computed values of the condensate fraction and chemical potential, as shown in the Appendices. In the top and middle panels, the strength is set to $\Lambda=10$.
In the top panel of Fig.~\ref{fig:Omega}, one can see that, as expected, the mean-field prediction of the Josephson frequency is correct at low-temperatures and low-gas parameters.
As denoted by the magenta line of the top panel, the mean-field theory is correct within a $1\%$-error margin when $\gamma\lesssim5\cdot 10^{-6}$ and $T^*\lesssim0.3$. There are then two distinct regions in which the role of the fluctuations cannot be neglected. The region for high temperature shows a significant decrease of the frequency, while for high gas parameter, the frequency is increased. Notice that the colormap extremes are fixed to $\pm 10\%$ for clarity, but larger corrections are expected for very high temperatures and very high gas parameters.
In those regions, one can use the asymptotic expressions obtained by expanding the condensate fraction and the chemical potential, whose validity regions are denoted by a black dashed line for high temperature, and a yellow dashed-dotted line for the low temperature.
Markers in the top panel correspond to parameters utilized in some experiments, detailed in Table~\ref{tab:comparison}.
\begin{table}[h]
\centering
\caption{Experimentally accessible regimes: gas parameter $\gamma$ and 
crossover temperature $T^\star$ in different bosonic Josephson junction experiments. The last column reports the marker used for each experiment in the top panel of Fig.~\ref{fig:Omega}.}
\label{tab:comparison}
\begin{tabular}{lcccc}
\hline\hline
Reference & specie & $\gamma$ & $T^\star$ & Marker \\

\hline
\vspace{-0.3cm} \\
Singh~\emph{et al.} (2025)\cite{singh2025}             & $^{87}$Rb & $3.77\times 10^{-5}$ & $0.103$ & \markSingh \\
Pezzè~\emph{et al.} (2024)\cite{pezze2024}             & $^6$Li$_2$  & $3.82\times 10^{-4}$ & $0.135$ & \markPezze \\
Bernhart~\emph{et al.} (2024)\cite{bernhart2024}       & $^{23}$Na & $3.65\times 10^{-5}$ & $0.160$ & \markBernhart \\
Albiez~\emph{et al.} (2005)\cite{albiez_direct_2005}   & $^{87}$Rb & $1.66\times 10^{-5}$ & $0.099$ & \markAlbiez \\
\hline\hline
\end{tabular}
\end{table}

At low temperature $T^*$ and high gas parameter $\gamma$, quantum fluctuations are stronger than thermal fluctuations; thus, the beyond-mean-field Josephson frequency is greater than the mean-field one. Conversely, when at low gas parameters $\gamma$ and higher temperatures $T^*$, thermal fluctuations dominate, and the Josephson frequency is lower than the mean-field prediction. When considering high values of both $T^*$ and $\gamma$, both thermal and quantum fluctuations are significant, as represented in the middle panels of Fig.~\ref{fig:Omega}; therefore, a slight variation in one of the two parameters results in an abrupt variation of the Josephson frequency.
Finally, in the bottom panel of Fig.~\ref{fig:Omega}, we illustrate how the corrections depend on the interaction strength $\Lambda$; specifically, we show that increasing $\Lambda$ increases the absolute deviation of the Josephson frequency from the mean-field prediction.

\section{Spontaneous symmetry breaking}
In the equations of motion, besides the stationary solution for initial conditions $(z_0,\phi_0)=(0,0)$, there exists a value of $\Lambda_s$ for which, for every $\Lambda>\Lambda_s$, there exist stationary solutions that break the $\mathbb{Z}_2$ site-swapping symmetry. Increasing $\Lambda$, the stationary points pass from the single value $(z_0,\phi_0)=(0,\pi)$ into the pair $(\pm z_s,\pi)$, effectively showing SSB.
The relation that defines the stationary points of Eq.~\ref{eq:smerzi-nostre} with $\phi = 0$ simplifies to considering the vanishing condition for the right hand side of the second equation. This gives the condition for the SSB critical strength $\Lambda_s$:
\begin{equation}
\lambda_{T}\bar\mu_{eq}^{1/2}\frac{z_0}{\sqrt{1-z^2_0}}=\Lambda_s\frac{\bar\mu_+-\bar\mu_-}{2}
\end{equation}

\begin{figure}[!htbp]
    \centering
    \includegraphics[width=\linewidth]{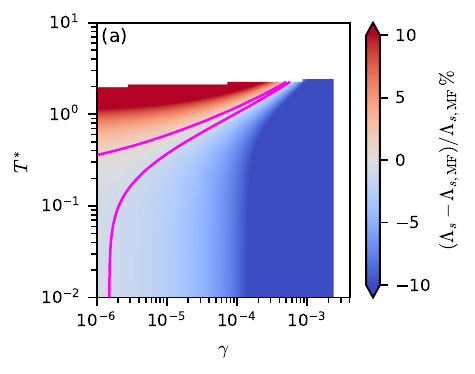}
    \includegraphics[width=\linewidth]{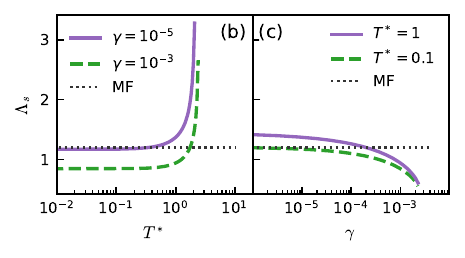}
    \includegraphics[width=\linewidth]{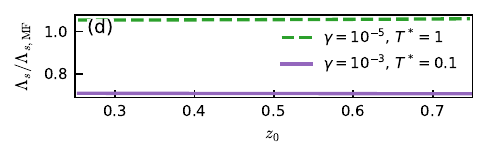}
    \caption{Corrections to the spontaneous symmetry breaking critical strength \ensuremath{\Lambda_s}. (a) Colormap of the relative beyond-mean-field correction of \ensuremath{\Lambda_s} for a bosonic gas at finite temperature \ensuremath{T^* =mk_BT/\hbar^2n^{2/3}}, gas parameter \ensuremath{\gamma=a_s^3n}, and initial total population imbalance \ensuremath{z_{tot,0}=0.55}. Line colors and markers follow the same convention as the top panel of Fig.~\ref{fig:Omega}. Middle panels show \ensuremath{\Lambda_s}: left panel (b) versus \ensuremath{T^*} at fixed \ensuremath{\gamma}, right panel (c) versus \ensuremath{\gamma} at fixed \ensuremath{T^*}. Panel (d) shows the dependence of the critical strength, normalized to the mean-field one, on the initial imbalance $z_0$.}
    \label{fig:LambdaSSB}
\end{figure}

Knowing the mean-field critical strength $\Lambda_{s,\text{MF}}=1/\sqrt{1-z^2_0}$ we now have that
\begin{equation}
    \frac{\Lambda_{s}}{\Lambda_{s,\text{MF}}}=\bar\mu_{eq}^{1/2}\frac{2\lambda_Tz_0}{\bar\mu_+-\bar\mu_-}
\end{equation}
which highlights the beyond-mean-field modification of the SSB threshold.
In Fig.~\ref{fig:LambdaSSB}, we show the resulting correction in the $(\gamma, T^*)$ plane (top panel), and as a function of $\gamma$ and $T^*$ only for selected parameters. The results, obtained with an initial population imbalance of $z_{tot, 0}=0.55$, show the competition effect between the thermal and quantum fluctuations. Their effect on the shift of $\Lambda_s$ is opposite to the one on the Josephson frequency: higher density tends to reduce $\Lambda_s$, whereas higher temperatures increase it. Moreover, the effect of density appears to be stronger than that of temperature in the parameter range of validity of the method. The dependence of the relative correction on the initial imbalance $z_0$ is very weak, as shown in the bottom panel of Fig.~\ref{fig:LambdaSSB}.

\section{Macroscopic quantum self-trapping}
\begin{figure}[!htbp]
    \centering
    \includegraphics[width=\linewidth]{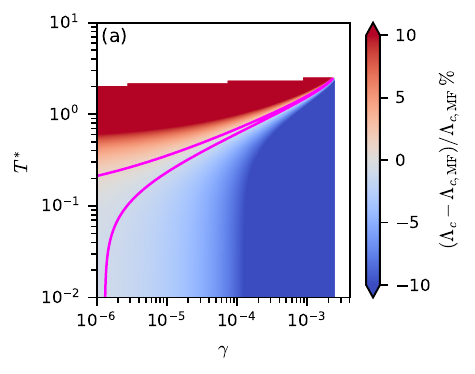}
    \includegraphics[width=\linewidth]{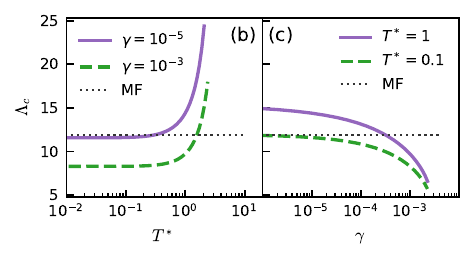}
        \includegraphics[width=\linewidth]{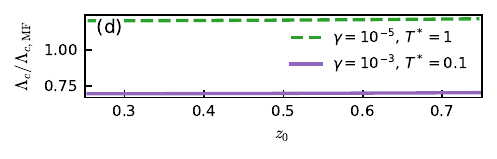}

    \caption{Corrections to the macroscopic quantum self-trapping critical strength \ensuremath{\Lambda_c}. (a) Colormap of the relative beyond-mean-field correction of \ensuremath{\Lambda_c} for a bosonic gas at finite temperature \ensuremath{T^* =mk_BT/\hbar^2n^{2/3}}, gas parameter \ensuremath{\gamma=a_s^3n}, and initial total population imbalance \ensuremath{z_{tot,0}=0.55}. Line colors and markers follow the same convention as the top panel of Fig.~\ref{fig:Omega}. Middle panels show \ensuremath{\Lambda_c}: left panel (b) versus \ensuremath{T^*} at fixed \ensuremath{\gamma}, right panel (c) versus \ensuremath{\gamma} at fixed \ensuremath{T^*}. Panel (d) shows the dependence of the critical strength, normalized to the mean-field one, on the initial imbalance $z_0$.}
    \label{fig:LambdaMQST}
\end{figure}

The conserved energy per unit of volume of the system is given by
\begin{align}
    \frac{E}{2L^3}=\frac{\mathcal{E}^++\mathcal{E^-}}{2}-Jn\sqrt{1-z^2}\cos{\phi}
\end{align}
where $\mathcal{E}^\pm\equiv\mathcal{E}(n^\pm,g,T)$ is the energy density of the two sites. To obtain the MQST condition, we impose that the conserved energy $E$ is greater than the energy of a system with initial condition $(z_0,\phi_0)=(0,\pi)$. The resulting inequality, dependent on the initial population imbalance $z_0$, allows one to define the MQST critical strength $\Lambda_c$, which is the strength for which, starting at $z_0$, the solutions are self-trapped and oscillate around a nonzero average population imbalance, namely
\begin{equation}
    \frac{\Lambda_{c}}{\Lambda_{c,\text{MF}}}=\bar\mu_{eq}^{1/2}\frac{\lambda_Tz^2_0}{\mathcal{E}^++\mathcal{E^-}-2\mathcal{E}}
\end{equation}
where $\mathcal{E}$ is the energy density of one site at equilibrium, and $\Lambda_{c,\text{MF}}$ is the mean-field critical strength.
In Fig.~\ref{fig:LambdaMQST} we show the relative correction in the $(\gamma, T^*)$ plane (top panel), as well as the dependence on $\gamma$ and $T^*$ for selected parameters (bottom panels). Results are obtained with $z_{tot, 0}=0.55$. The correction behaviour for $\Lambda_c$ is similar to that of the SSB critical strength $\Lambda_s$. However, the effect of the temperature is stronger for a similar choice of parameters. Similarly to the SSB case, also in this case the dependence of the relative correction on the imbalance $z_0$ is very weak.

\section{Conclusions}
In this work, we have developed an extension of the two-mode description of a bosonic Josephson junction that incorporates both quantum and thermal fluctuations in three spatial dimensions. This approach relies on approximating the fluctuations in each well with those of a homogeneous Bose gas, allowing us to retain a tractable two-site model while including beyond-mean-field corrections in a semi-analytical form.

Within this framework, we derived modified expressions for the Josephson frequency, the critical strength for macroscopic quantum self-trapping, and the threshold for spontaneous symmetry breaking. While reproducing previous results for zero-temperature Gaussian fluctuations \cite{bardin_quantum_2024}, our approach naturally extends them to finite temperature. All these quantities exhibit a competition between quantum and thermal contributions. Specifically, thermal fluctuations lead to a reduction of the Josephson frequency and to an increase of both critical strengths, while quantum fluctuations act in the opposite direction. This opposite behavior provides a consistent picture of how different sources of fluctuations affect the nonlinear dynamics of the junction.

The analysis of experimentally relevant parameters indicates that, in typical realizations of bosonic Josephson junctions, the system operates in a regime where quantum fluctuations dominate over thermal ones, although the latter can become relevant at sufficiently high temperatures or low densities. The present results therefore quantify the range of validity of mean-field predictions and provide corrections that can be directly compared with experiments.

The present model is based on the assumption of a stationary non-condensed fraction and neglects its transport across the junction. Extending the formalism to include the dynamics of the thermal cloud and its coupling to the condensate would allow one to investigate dissipative effects and relaxation processes, and it represents a natural direction for future work.

\section*{Acknowledgments} 
LS acknowledges the BIRD Project “Ultra-cold atoms in curved geometries” of the University of Padova. LS is partially supported by the European Union-NextGenerationEU within the National Center for HPC, Big Data, and Quantum Computing [Project No. CN00000013, CN1 Spoke 10: Quantum Computing] and by the European Quantum Flagship Project PASQuanS 2. LS acknowledges Iniziativa Specifica Quantum of Istituto Nazionale di Fisica Nucleare, the Project “Frontiere Quantistiche” within the 2023 funding programme ‘Dipartimenti di Eccellenza’ of the Italian Ministry for Universities and Research, and the PRIN 11 2022 Project “Quantum Atomic Mixtures: Droplets, Topological Structures, and Vortices”.

\appendix
\section{Chemical Potential of a uniform bosonic gas}
\label{Appendix A}
The chemical potential of a uniform interacting bosonic gas of density $n=N/L^3$ at temperature $T$~\cite{salasnich_zero-point_2016} can be obtained starting from its beyond-mean-field Grand Potential, which accounts for quantum and thermal fluctuations, namely:
\begin{equation}
    \frac{\Omega}{L^3}=-\frac{\mu^2}{2g}+\frac{8\mu^{5/2}}{15\pi^2}\sqrt{\frac{m}{\hbar^2}}^3\left[1-\frac{5}{2}\int_0^\infty dt\, 
\frac{\sinh^3(t/2)\cosh(t)}{e^{\beta\mu\sinh t} - 1}\right]
\end{equation}
Deriving the opposite of the grand potential with respect to the chemical potential, we obtain the number density of the system and thus an implicit formula for the chemical potential, namely
\begin{equation}
    n=\frac{\mu}{g}\left\{1-\frac{4\mu^{1/2}g}{3\pi^2}\sqrt{\frac{m}{\hbar^2}}^3\left[1+\frac{3}{2}I(\beta\mu)\right]\right\}
\end{equation}
where $I(\beta\mu)$ is the following integral
\begin{equation}
    I(\beta\mu)=\int_0^\infty dt\, 
\frac{\sinh^3(t/2)}{e^{\beta\mu\sinh t} - 1}
\end{equation}
Since we are interested in the relative correction of the mean-field chemical potential, we now define the renormalization of the chemical potential over its mean-field value as $\bar\mu=\mu/gn$. We obtain an implicit equation for $\bar\mu$ only in terms of the gas parameter $\gamma\equiv a_s^3n$ and  an integral parameter $\alpha\equiv\beta gn=4\pi\gamma^{1/3}/T^*$, which depends on the dimensionless temperature $T^*\equiv m k_B T / \hbar^2 n^{3/2}$. The implicit equation for the renormalized chemical potential is 
\begin{equation}
    1=\bar\mu\left\{1-\frac{32\sqrt\gamma}{3\sqrt\pi}\bar\mu^{1/2}\left[1+\frac{3}{2}I(\alpha\bar\mu)\right]\right\}
\end{equation}
Since a bosonic Josephson junction consists of two uniform subsystems with initial number densities 
$n^\pm = n(1 \pm z_{\text{tot},0})$, 
it is convenient to express these as dilations of the equilibrium density $n$ by the factors 
$1 \pm z_{\text{tot},0}$. 
In general, when the density is dilated by a factor $x$, i.e. $n \rightarrow n x$, the dimensionless parameters scale as
\begin{equation}
 \gamma \rightarrow \gamma x, 
 \qquad   
 T^* \rightarrow T^* x^{-2/3},\qquad \bar\mu\rightarrow \bar\mu x^{-1}.
\end{equation}
To find the chemical potential of a subsystem with number density $nx$, it is convenient to define an implicit relation for the renormalized chemical potential $\bar\mu$, obtaining
\begin{equation}
    x=\bar\mu\left\{1-\frac{32\sqrt\gamma}{3\sqrt\pi}\bar\mu^{1/2}\left[1+\frac{3}{2}I(\alpha\bar\mu)\right]\right\}
\end{equation}
which does not depend on $x$ on the right-hand side since $\gamma$ and $\alpha$ scale linearly with $x$. 
Note that at the mean-field level, one obtains $\bar\mu=x$, as expected, since in such case the chemical potential is linear in the number density.
Finally, the chemical potential at the two sites $\mu_\pm$ is found by implicitly solving
\begin{equation}
    1\pm z_{tot}=\bar\mu_\pm\left\{1-\frac{32\sqrt\gamma}{3\sqrt\pi}\bar\mu_\pm^{1/2}\left[1+\frac{3}{2}I(\alpha\bar\mu_\pm)\right]\right\}
\end{equation}
\begin{figure}[h]
    \centering
    \includegraphics[width=\linewidth]{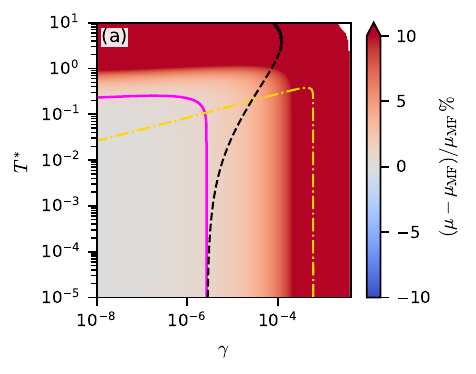}
    \includegraphics[width=\linewidth]{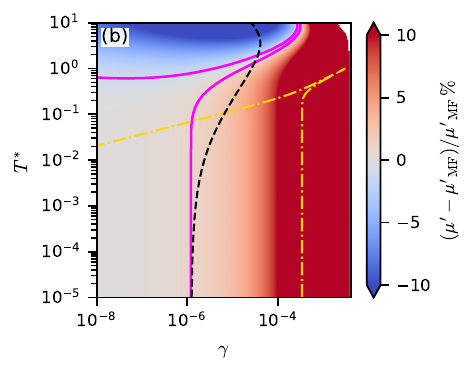}
    \caption{Colormap of the relative beyond-mean-field  correction of the chemical potential (a) and chemical potential derivative (b) for a bosonic gas at finite temperature $T^* =mk_BT/\hbar^2n^{2/3}$ and gas parameter $\gamma=a_s^3n$. In both plots the bold magenta line represent relative corrections of $1\%$, while the black dashed line and the yellow dashed-dotted line represent the limit above which the high-temperature approximation and the low-temperature approximation are not valid anymore.}
    \label{fig:mu,muprime}
\end{figure}
We are now interested in the two limiting regions of the integral $I(\alpha\bar\mu)$, which we refer to as the high-temperature regime  $\alpha\bar\mu\ll1$, where the integral is divergent as  $\alpha\bar\mu\to$, and the low-temperature regime $\alpha\bar\mu\gg1$, in which the integral decays to zero. To do so, we need to use Mellin transforms.
\subsection{High Temperature Limit}
In the limit of $\alpha\bar\mu\ll1$,  the integral $I(\alpha \bar{\mu})$ can be expanded as the convergent series

\begin{equation}
\begin{split}
    I(\alpha\bar\mu)=&-\frac{\pi}{2\alpha\bar\mu}-\frac{2}{3}+\sum_{k=0}^\infty I_k(\alpha\bar\mu)^{-3/2+k}
\end{split}
\end{equation}
    with coefficients $I_k=(-1)^{\lfloor (k+1)/2 \rfloor}\frac{\sqrt{\pi}(2k+1)}{4\sqrt{2}k!}\zeta(3/2-k)$. The series converges for  $\alpha\bar\mu\leq2\pi$.
    Note that the constant term is such that it cancels out precisely the term that arises from the quantum fluctuations. Considering only the leading term, we have the familiar equation
    \begin{equation}
    \bar\mu=x+\left(\frac{T^{*}}{T^*_c}\right)^{3/2}
    \end{equation}
    where in dimensionless units the BEC critical temperature is just $T^*_c=\zeta(3/2)^{-2/3}\simeq3.3125$.
    If we instead consider the first four terms, we have a quadratic equation in $\bar\mu^{1/2}$ of the form $A\bar\mu+2B\bar\mu^{1/2}-C=0$, where
\begin{subequations}
\begin{align}
A&=1+3\gamma^{1/3}\zeta(1/2)\sqrt{\frac{2T^*}{\pi}}\\
    B&=\frac{\left(T^*\right)\gamma^{1/6}}{\sqrt{\pi}}\\
    C&=x+\frac{\left(T^*\right)^{3/2}}{\left(T^*_c\right)^{3/2}}
\end{align}
\end{subequations}
and its solution is
\begin{equation}
\begin{split}\bar\mu=&\frac{C}{A}+\frac{2B^2}{A^2}\left(1-\sqrt{1+\frac{AC}{B^2}}\right)
    \end{split}
\end{equation}
We have then that its derivative with respect to $x$ is
\begin{equation}
    \frac{d\bar\mu}{dx}=\frac{1-\sqrt{1+AC/B^2}^{-1}}{A}
\end{equation}

\subsection{Low-Temperature Limit}
In the low-temperature limit, where $\alpha\bar\mu\gg1$, the integral $I(\alpha\bar\mu)$ admits an asymptotic expansion of the form
\begin{equation}
    I(\alpha\bar\mu)=\sum_{k=2}^{\infty}\tilde{I}_k(\alpha\bar\mu)^{-2k}
\end{equation}
with coefficients $\tilde{I}_k=(-1)^{k}\frac{\zeta(2k)\Gamma(2k-3/2)(k-1)}{\sqrt\pi}\alpha^{-2k}$.This series is asymptotic rather than convergent: adding successive terms improves the approximation only up to an optimal order that depends on $\alpha\bar\mu$. Consequently, the truncated series becomes increasingly accurate for larger values of $\alpha\bar\mu$, but including too many terms leads to reduced accuracy when $\alpha\bar\mu$ is smaller. For this reason, in the following, we will consider only the first two terms of the expansion in the low-temperature limit, namely
\begin{equation}
    x=\bar\mu-\frac{32\sqrt\gamma}{3\sqrt\pi}\bar\mu^{3/2}-\frac{T^{*4}}{2^{7}15\sqrt\pi\gamma^{5/6}}\bar\mu^{-5/2}+\frac{T^{*6}}{2^{11}9\sqrt\pi\gamma^{3/2}}\bar\mu^{-9/2}
\end{equation}
It is then possible to perturbatively invert the above relation and obtain $\bar\mu$ as a function of the dilation $x$. At the first perturbation order the chemical potential is given by
\begin{equation}
    \mu=x+\frac{32\sqrt\gamma}{3\sqrt\pi}x^{3/2}+\frac{T^{*4}}{2^{7}15\sqrt\pi\gamma^{5/6}}x^{-5/2}-\frac{T^{*6}}{2^{11}9\sqrt\pi\gamma^{3/2}}x^{-9/2}
\end{equation}
while the second perturbation order expression is
\begin{equation}
\begin{aligned}
\mu = x
&+ \Biggl[
    \frac{32\sqrt{\gamma}}{3\sqrt{\pi}} x^{3/2}
    + \frac{T^{*4}}{2^{7} 15 \sqrt{\pi}\, \gamma^{5/6}} x^{-5/2}
\\
&\qquad
    - \frac{T^{*6}}{2^{11} 9 \sqrt{\pi}\, \gamma^{3/2}} x^{-9/2}
\Biggr]
\\
&\quad\times \Biggl[
    1
    + \frac{16\sqrt{\gamma}}{\sqrt{\pi}} x^{1/2}
\\
&\qquad
    - \frac{T^{*4}}{2^{8} 3 \sqrt{\pi}\, \gamma^{5/6}} x^{-7/2}
    + \frac{T^{*6}}{2^{12} \sqrt{\pi}\, \gamma^{3/2}} x^{-11/2}
\Biggr].
\end{aligned}
\end{equation}
which provides the second-order low-temperature correction to the chemical potential as a function of the dilation $x$.

\section{Condensate Fraction of a uniform bosonic gas}
\label{Appendix B}
For a uniform bosonic gas of number density $n = N / L^3$ at temperature $T$ with contact interaction coupling $g$, the condensate fraction $\lambda$ is given by the following self-consistent formula~\cite{Vianello2024-ed}:
\begin{equation}
    \lambda(\gamma, T^*) =
    1 - \frac{8\sqrt{\gamma}}{3\sqrt{\pi}}
    \left[
        1 + 3Y(\alpha)
    \right] \,,
    \label{CondFrac}
\end{equation}
which depends on the gas parameter $\gamma$, and on the dimensionless temperature
$T^*$ through the parameter $\alpha = 4\pi \gamma^{1/3} / T^* = \beta g n$  entering the following integral
\begin{equation}
Y(\alpha) \equiv 
\int_0^\infty dt\, 
\frac{\sinh(t/2)\cosh(t)}{e^{\alpha\sinh t} - 1}.
\end{equation}
which accounts for the thermal depletion of the condensate. 
This expression is obtained by a change of variables applied to the 
two-parameter integral appearing in Ref.~\cite{Vianello2024-ed}.
As stated in Appendix \ref{Appendix A}, since our system consists of two uniform subsystems with initial number densities 
$n^\pm = n(1 \pm z_{\text{tot}})$, 
it is convenient to work in terms of the dilation of the number density by a factor $x$, i.e., $n \rightarrow n x$, which results in a corresponding condensate fraction given by
\begin{equation}
    \lambda(x, \gamma, T^*) =
    1 - \frac{8\sqrt{\gamma}}{3\sqrt{\pi}} x^{1/2}
    \left[1 + 3Y(\alpha x)\right].
\end{equation}
We then set the initial condensate fractions of the two subsystems as $\lambda_0^\pm = \lambda(1 \pm z_{\text{tot},0}, \gamma, T^*)$. \\
Similarly to what was done for the integral $I(\alpha\bar\mu)$ inside the chemical potential implicit equation in App.\ref{Appendix A}, we are interested in the behavior of the integral in the high- and low-temperature limits. Using Mellin transforms, we see that the coefficients of the series expansion of the two integrals are connected. Note that while the integral parameter in App.\ref{Appendix A} was $\alpha\bar\mu$, for the condensate fraction it is just $\alpha x$, and this corresponds to the mean-field value of $\alpha\bar\mu$.
\begin{figure}[h]
    \centering
    \includegraphics[width=\linewidth]{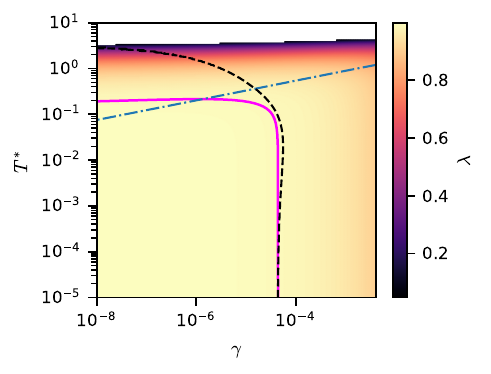}
    \caption{Colormap of the condensate fraction $\lambda$ for a bosonic gas at finite temperature $T^* =mk_BT/\hbar^2n^{2/3}$ and gas parameter $\gamma=a_s^3n$. The bold magenta line represent relative corrections of $1\%$, while the black dashed line and the blue dashed-dotted line represent the limit above which the high-temperature approximation and the low-temperature approximation are not valid.}
    \label{fig:lambda}
\end{figure}
\subsection{High-Temperature limit}
In the limit of $\alpha x\ll1$, the integral $Y(\alpha x)$ can be expanded as the convergent series 
\begin{equation}
    Y(\alpha x)=-\frac{\pi}{2\alpha x}-\frac{1}{3}+\sum_{k=0}^\infty Y_k(\alpha x)^{-3/2+k} \,,
\end{equation}
with coefficients $Y_k=(-1)^{\lfloor (k+1)/2 \rfloor}\frac{\sqrt{\pi}}{2\sqrt{2}k!}\zeta(3/2-k)$. The series converges for $\alpha\bar\mu\leq2\pi$, exactly as does the corresponding series for the integral entering the chemical potential. This is not surprising, since the coefficients $I_k$ are directly related to the coefficients $Y_k$of that expansion through
\begin{equation}
    I_k=\frac{2k+1}{2}Y_k \,.
\end{equation}
Note that, also in this case, the constant term in the expansion is such that it cancels out the quantum fluctuation contribution in the condensate fraction expression.
If $\alpha x<2\pi$, the condensate fraction can be written exactly as the series
\begin{equation}
    \lambda(x,\gamma,T^*)=1-\frac{T^{*3/2}}{T_{c,x}^{*3/2}}+T^*\sqrt{\frac{\gamma^{1/3}}{x\pi}}-\frac{T^{*3/2}}{x\pi^2 }\sum_{k=1}^\infty Y_k(\alpha x)^{k}
\end{equation}
where $T_{c,x}^*\equiv2\pi x^{2/3}/\zeta(3/2)^{2/3}$ is the mean-field BEC critical temperature.
The off-set $\tilde{z}_0$ in case of infinitesimal imbalance $z_{tot}\ll1$ is
\begin{equation}
    \begin{split}
        \tilde{z}_0\frac{\lambda_{eq}}{z_{tot,0}}&=\lambda_{eq}+\lambda'_{eq}-1\\
        &=\frac{T^*}{2}\sqrt{\frac{\gamma^{1/3}}{\pi}}-\frac{T^{*3/2}}{\pi^2}\sum_{k=1}^\infty kY_k\alpha^{k} \,.\\
    \end{split}
\end{equation}
\subsection{Low-Temperature limit}
In the low-temperature limit, where $\alpha x\gg1$, the integral $Y(\alpha x)$ has an asymptotic expansion of the form 
\begin{equation}
    Y(\alpha x)=\sum_{k=1}^{\infty}\tilde{Y}_k(\alpha x)^{-2k} \,,
\end{equation}
with coefficients $\tilde{Y}_k=(-1)^{k+1}\zeta(2k)\Gamma(3/2-k)/2\sqrt{\pi}$ that are related to the coefficients $\tilde{I}_k$ by the relation
\begin{equation}
    \tilde I_k=2(1-k)\tilde Y_k \,.
\end{equation}
Since the series is only asymptotic, higher-order terms provide fine tuning corrections to the condensate fraction for higher values of $\alpha x$. We will therefore consider just the first two orders in the expansion of the condensate fraction in the low-temperature limit, namely
\begin{equation}
\begin{split}
    \lambda(x,\gamma,T^*)=&1-\frac{8\sqrt\gamma}{3\sqrt{\pi}}x^{1/2}+\\
    &-\frac{T^{*2}}{24\sqrt{\pi}\gamma^{1/6}}x^{-3/2}+\frac{T^{*4}}{7680\sqrt{\pi}\gamma^{5/6}}x^{-7/2} \,,
\end{split}
\end{equation}
which gives the leading low-temperature correction to the condensate fraction.

\bibliography{ajj_merged}

@article{josephson_possible_1962,
  author = {Josephson, B. D.},
  doi = {10.1016/0031-9163(62)91369-0},
  issn = {0031-9163},
  journal = {Phys. Lett.},
  month = {July},
  number = {7},
  pages = {251--253},
  title = {Possible new effects in superconductive tunnelling},
  url = {https://www.sciencedirect.com/science/article/pii/0031916362913690},
  urldate = {2023-11-06},
  volume = {1},
  year = {1962}
}

@article{smerzi_quantum_1997,
  author = {Smerzi, A. and Fantoni, S. and Giovanazzi, S. and Shenoy, S. R.},
  doi = {10.1103/PhysRevLett.79.4950},
  journal = {Phys. Rev. Lett.},
  month = {December},
  number = {25},
  pages = {4950--4953},
  title = {Quantum {Coherent} {Atomic} {Tunneling} between {Two} {Trapped} {Bose}-{Einstein} {Condensates}},
  url = {https://link.aps.org/doi/10.1103/PhysRevLett.79.4950},
  urldate = {2023-11-06},
  volume = {79},
  year = {1997}
}

@article{albiez_direct_2005,
  author = {Albiez, Michael and Gati, Rudolf and Fölling, Jonas and Hunsmann, Stefan and Cristiani, Matteo and Oberthaler, Markus K.},
  doi = {10.1103/PhysRevLett.95.010402},
  issn = {0031-9007},
  journal = {Phys. Rev. Lett.},
  month = {July},
  number = {1},
  pages = {010402},
  pmid = {16090588},
  title = {Direct observation of tunneling and nonlinear self-trapping in a single bosonic {Josephson} junction},
  volume = {95},
  year = {2005}
}

@article{valtolina_josephson_2015,
  author = {Valtolina, Giacomo and Burchianti, Alessia and Amico, Andrea and Neri, Elettra and Xhani, Klejdja and Seman, Jorge Amin and Trombettoni, Andrea and Smerzi, Augusto and Zaccanti, Matteo and Inguscio, Massimo and Roati, Giacomo},
  doi = {10.1126/science.aac9725},
  issn = {1095-9203},
  journal = {Science},
  month = {December},
  number = {6267},
  pages = {1505--1508},
  pmid = {26680193},
  title = {Josephson effect in fermionic superfluids across the {BEC}-{BCS} crossover},
  volume = {350},
  year = {2015}
}

@article{spuntarelli_josephson_2007,
  author = {Spuntarelli, A. and Pieri, P. and Strinati, G. C.},
  doi = {10.1103/PhysRevLett.99.040401},
  journal = {Phys. Rev. Lett.},
  month = {July},
  number = {4},
  pages = {040401},
  title = {Josephson {Effect} throughout the {BCS}-{BEC} {Crossover}},
  url = {https://link.aps.org/doi/10.1103/PhysRevLett.99.040401},
  urldate = {2023-11-06},
  volume = {99},
  year = {2007}
}

@article{cataliotti_josephson_2001,
  author = {Cataliotti, F. S. and Burger, S. and Fort, C. and Maddaloni, P. and Minardi, F. and Trombettoni, A. and Smerzi, A. and Inguscio, M.},
  doi = {10.1126/science.1062612},
  issn = {0036-8075},
  journal = {Science},
  month = {August},
  number = {5531},
  pages = {843--846},
  pmid = {11486083},
  title = {Josephson junction arrays with {Bose}-{Einstein} condensates},
  volume = {293},
  year = {2001}
}

@article{tian_quantum_2003,
  author = {Tian, Lin and Zoller, P.},
  doi = {10.1103/PhysRevA.68.042321},
  journal = {Phys. Rev. A},
  month = {October},
  number = {4},
  pages = {042321},
  title = {Quantum computing with atomic {Josephson} junction arrays},
  url = {https://link.aps.org/doi/10.1103/PhysRevA.68.042321},
  urldate = {2023-11-06},
  volume = {68},
  year = {2003}
}

@article{leggett_concept_1991,
  author = {Leggett, Anthony J. and Sols, Fernando},
  doi = {10.1007/BF01883640},
  issn = {1572-9516},
  journal = {Found. Phys.},
  month = {March},
  number = {3},
  pages = {353--364},
  title = {On the concept of spontaneously broken gauge symmetry in condensed matter physics},
  url = {https://doi.org/10.1007/BF01883640},
  urldate = {2023-11-06},
  volume = {21},
  year = {1991}
}

@article{wimberger_finite-size_2021,
  author = {Wimberger, Sandro and Manganelli, Gabriele and Brollo, Alberto and Salasnich, Luca},
  doi = {10.1103/PhysRevA.103.023326},
  journal = {Phys. Rev. A},
  month = {February},
  number = {2},
  pages = {023326},
  title = {Finite-size effects in a bosonic {Josephson} junction},
  url = {https://link.aps.org/doi/10.1103/PhysRevA.103.023326},
  urldate = {2023-11-06},
  volume = {103},
  year = {2021}
}

@article{salasnich_zero-point_2016,
  author = {Salasnich, Luca and Toigo, Flavio},
  doi = {10.1016/j.physrep.2016.06.003},
  issn = {0370-1573},
  journal = {Phys. Rep.},
  month = {June},
  pages = {1--29},
  title = {Zero-point energy of ultracold atoms},
  url = {https://www.sciencedirect.com/science/article/pii/S0370157316301338},
  urldate = {2023-11-06},
  volume = {640},
  year = {2016}
}

@article{levy_c_2007,
  author = {Levy, S. and Lahoud, E. and Shomroni, I. and Steinhauer, J.},
  doi = {10.1038/nature06186},
  issn = {1476-4687},
  journal = {Nature},
  month = {October},
  number = {7162},
  pages = {579--583},
  pmid = {17914391},
  title = {The a.c. and d.c. {Josephson} effects in a {Bose}-{Einstein} condensate},
  volume = {449},
  year = {2007}
}

@article{raghavan_coherent_1999,
  author = {Raghavan, S. and Smerzi, A. and Fantoni, S. and Shenoy, S. R.},
  doi = {10.1103/PhysRevA.59.620},
  journal = {Phys. Rev. A},
  month = {January},
  number = {1},
  pages = {620--633},
  shorttitle = {Coherent oscillations between two weakly coupled {Bose}-{Einstein} condensates},
  title = {Coherent oscillations between two weakly coupled {Bose}-{Einstein} condensates: {Josephson} effects, $\pi$ oscillations, and macroscopic quantum self-trapping},
  url = {https://link.aps.org/doi/10.1103/PhysRevA.59.620},
  urldate = {2023-11-06},
  volume = {59},
  year = {1999}
}

@article{xhani_dynamical_2020,
  author = {Xhani, K and Galantucci, L and Barenghi, C F and Roati, G and Trombettoni, A and Proukakis, N P},
  doi = {10.1088/1367-2630/abc8e4},
  issn = {1367-2630},
  journal = {New J. Phys.},
  month = {December},
  number = {12},
  pages = {123006},
  title = {Dynamical phase diagram of ultracold {Josephson} junctions},
  url = {https://dx.doi.org/10.1088/1367-2630/abc8e4},
  urldate = {2025-03-10},
  volume = {22},
  year = {2020}
}

@article{zaccantizwerger,
  author = {Zaccanti, M. and Zwerger, W.},
  doi = {10.1103/PhysRevA.100.063601},
  issue = {6},
  journal = {Phys. Rev. A},
  month = {December},
  number = {6},
  numpages = {8},
  pages = {063601},
  publisher = {American Physical Society},
  title = {Critical {Josephson} current in {BCS}-{BEC}--crossover superfluids},
  url = {https://link.aps.org/doi/10.1103/PhysRevA.100.063601},
  urldate = {2023-11-06},
  volume = {100},
  year = {2019}
}

@article{del_pace_tunneling_2021,
  author = {Del Pace, G. and Kwon, W. J. and Zaccanti, M. and Roati, G. and Scazza, F.},
  doi = {10.1103/PhysRevLett.126.055301},
  journal = {Phys. Rev. Lett.},
  month = {February},
  number = {5},
  pages = {055301},
  title = {Tunneling {Transport} of {Unitary} {Fermions} across the {Superfluid} {Transition}},
  url = {https://link.aps.org/doi/10.1103/PhysRevLett.126.055301},
  urldate = {2023-11-06},
  volume = {126},
  year = {2021}
}

@article{kwon_strongly_2020,
  author = {Kwon, W. J. and Del Pace, G. and Panza, R. and Inguscio, M. and Zwerger, W. and Zaccanti, M. and Scazza, F. and Roati, G.},
  doi = {10.1126/science.aaz2463},
  journal = {Science},
  month = {July},
  number = {6499},
  pages = {84--88},
  title = {Strongly correlated superfluid order parameters from dc {Josephson} supercurrents},
  url = {https://www.science.org/doi/10.1126/science.aaz2463},
  urldate = {2023-11-06},
  volume = {369},
  year = {2020}
}

@article{piselli_josephson_2020,
  author = {Piselli, V. and Simonucci, S. and Strinati, G. Calvanese},
  date = {2020-10-19},
  doi = {10.1103/PhysRevB.102.144517},
  journal = {Phys. Rev. B},
  number = {14},
  pages = {144517},
  title = {Josephson effect at finite temperature along the {BCS}-{BEC} crossover},
  url = {https://link.aps.org/doi/10.1103/PhysRevB.102.144517},
  urldate = {2025-02-07},
  volume = {102},
  year={2020}
}

@article{xhani_dissipation_2022,
  author = {Xhani, K. and Proukakis, N. P.},
  date = {2022-09-13},
  doi = {10.1103/PhysRevResearch.4.033205},
  journal = {Phys. Rev. Res.},
  number = {3},
  pages = {033205},
  title = {Dissipation in a finite-temperature atomic Josephson junction},
  url = {https://link.aps.org/doi/10.1103/PhysRevResearch.4.033205},
  urldate = {2025-02-07},
  volume = {4},
  year={2022}
}

@article{bidasyuk_finite-temperature_2018,
  author = {Bidasyuk, Y M and Weyrauch, M and Momme, M and Prikhodko, O O},
  date = {2018-09},
  doi = {10.1088/1361-6455/aae022},
  issn = {0953-4075},
  journal = {J. Phys. B: At. Mol. Opt. Phys.},
  langid = {english},
  number = {20},
  year={2018},
  pages = {205301},
  title = {Finite-temperature dynamics of a bosonic Josephson junction},
  url = {https://dx.doi.org/10.1088/1361-6455/aae022},
  urldate = {2025-02-07},
  volume = {51}
}

@article{burchianti_connecting_2018,
  author = {Burchianti, A. and Scazza, F. and Amico, A. and Valtolina, G. and Seman, J. A. and Fort, C. and Zaccanti, M. and Inguscio, M. and Roati, G.},
  doi = {10.1103/PhysRevLett.120.025302},
  journal = {Phys. Rev. Lett.},
  month = {January},
  number = {2},
  pages = {025302},
  title = {Connecting {Dissipation} and {Phase} {Slips} in a {Josephson} {Junction} between {Fermionic} {Superfluids}},
  url = {https://link.aps.org/doi/10.1103/PhysRevLett.120.025302},
  urldate = {2025-03-10},
  volume = {120},
  year = {2018}
}

@article{wlazlowski_dissipation_2023,
  author = {Wlazłowski, Gabriel and Xhani, Klejdja and Tylutki, Marek and Proukakis, Nikolaos P. and Magierski, Piotr},
  doi = {10.1103/PhysRevLett.130.023003},
  journal = {Phys. Rev. Lett.},
  month = {January},
  number = {2},
  pages = {023003},
  title = {Dissipation {Mechanisms} in {Fermionic} {Josephson} {Junction}},
  url = {https://link.aps.org/doi/10.1103/PhysRevLett.130.023003},
  urldate = {2025-03-10},
  volume = {130},
  year = {2023}
}

@article{avenel_josephson_1988,
  author = {Avenel, O. and Varoquaux, E.},
  doi = {10.1103/PhysRevLett.60.416},
  journal = {Phys. Rev. Lett.},
  month = {February},
  number = {5},
  pages = {416--419},
  title = {Josephson effect and quantum phase slippage in superfluids},
  url = {https://link.aps.org/doi/10.1103/PhysRevLett.60.416},
  urldate = {2025-03-10},
  volume = {60},
  year = {1988}
}

@article{bardin_quantum_2024,
  author = {Bardin, A and Lorenzi, F and Salasnich, L},
  doi = {10.1088/1367-2630/ad127b},
  issn = {1367-2630},
  journal = {New J. Phys.},
  month = {January},
  number = {1},
  pages = {013021},
  shorttitle = {Quantum fluctuations in atomic {Josephson} junctions},
  title = {Quantum fluctuations in atomic {Josephson} junctions: the role of dimensionality},
  url = {https://dx.doi.org/10.1088/1367-2630/ad127b},
  urldate = {2025-03-10},
  volume = {26},
  year = {2024}
}

@article{furutani_interaction-induced_2024,
  author = {Furutani, Koichiro and Salasnich, Luca},
  doi = {10.1103/PhysRevB.110.L140503},
  journal = {Phys. Rev. B},
  month = {October},
  number = {14},
  pages = {L140503},
  title = {Interaction-induced dissipative quantum phase transition in a head-to-tail atomic {Josephson} junction},
  url = {https://link.aps.org/doi/10.1103/PhysRevB.110.L140503},
  urldate = {2025-03-10},
  volume = {110},
  year = {2024}
}

@article{furutani_quantum_2021,
  author = {Furutani, Koichiro and Salasnich, Luca},
  doi = {10.1103/PhysRevB.104.014519},
  journal = {Phys. Rev. B},
  month = {July},
  number = {1},
  pages = {014519},
  title = {Quantum and thermal fluctuations in the dynamics of a resistively and capacitively shunted {Josephson} junction},
  url = {https://link.aps.org/doi/10.1103/PhysRevB.104.014519},
  urldate = {2025-03-10},
  volume = {104},
  year = {2021}
}

@article{burchianti_josephson_2017,
  author = {Burchianti, Alessia and Fort, Chiara and Modugno, Michele},
  doi = {10.1103/PhysRevA.95.023627},
  journal = {Phys. Rev. A},
  month = {February},
  number = {2},
  pages = {023627},
  shorttitle = {Josephson plasma oscillations and the {Gross}-{Pitaevskii} equation},
  title = {Josephson plasma oscillations and the {Gross}-{Pitaevskii} equation: {Bogoliubov} approach versus two-mode model},
  url = {https://link.aps.org/doi/10.1103/PhysRevA.95.023627},
  urldate = {2025-03-10},
  volume = {95},
  year = {2017}
}

@article{anderson_probable_1963,
  author = {Anderson, P. W. and Rowell, J. M.},
  doi = {10.1103/PhysRevLett.10.230},
  journal = {Phys. Rev. Lett.},
  month = {March},
  number = {6},
  pages = {230--232},
  title = {Probable {Observation} of the {Josephson} {Superconducting} {Tunneling} {Effect}},
  url = {https://link.aps.org/doi/10.1103/PhysRevLett.10.230},
  urldate = {2025-03-10},
  volume = {10},
  year = {1963}
}

@article{amico_roadmap_2021,
  author = {Amico, L. and Boshier, M. and Birkl, G. and Minguzzi, A. and Miniatura, C. and Kwek, L.-C. and Aghamalyan et al.},
  doi = {10.1116/5.0026178},
  issn = {2639-0213},
  journal = {AVS Quantum Science},
  month = {August},
  number = {3},
  pages = {039201},
  shorttitle = {Roadmap on {Atomtronics}},
  title = {Roadmap on {Atomtronics}: {State} of the art and perspective},
  url = {https://doi.org/10.1116/5.0026178},
  urldate = {2025-03-10},
  volume = {3},
  year = {2021}
}

@article{amico_colloquium_2022,
  author = {Amico, Luigi and Anderson, Dana and Boshier, Malcolm and Brantut, Jean-Philippe and Kwek, Leong-Chuan and Minguzzi, Anna and von Klitzing, Wolf},
  doi = {10.1103/RevModPhys.94.041001},
  journal = {Rev. Mod. Phys.},
  month = {November},
  number = {4},
  pages = {041001},
  shorttitle = {Colloquium},
  title = {Colloquium: {Atomtronic} circuits: {From} many-body physics to quantum technologies},
  url = {https://link.aps.org/doi/10.1103/RevModPhys.94.041001},
  urldate = {2025-03-10},
  volume = {94},
  year = {2022}
}

@article{shin_atom_2004,
  author = {Shin, Y. and Saba, M. and Pasquini, T. A. and Ketterle, W. and Pritchard, D. E. and Leanhardt, A. E.},
  doi = {10.1103/PhysRevLett.92.050405},
  journal = {Phys. Rev. Lett.},
  month = {February},
  number = {5},
  pages = {050405},
  title = {Atom {Interferometry} with {Bose}-{Einstein} {Condensates} in a {Double}-{Well} {Potential}},
  url = {https://link.aps.org/doi/10.1103/PhysRevLett.92.050405},
  urldate = {2025-03-10},
  volume = {92},
  year = {2004}
}

@article{anderson_macroscopic_1998,
  author = {Anderson, B. P. and Kasevich, M. A.},
  doi = {10.1126/science.282.5394.1686},
  journal = {Science},
  month = {November},
  number = {5394},
  pages = {1686--1689},
  title = {Macroscopic {Quantum} {Interference} from {Atomic} {Tunnel} {Arrays}},
  url = {https://www.science.org/doi/full/10.1126/science.282.5394.1686},
  urldate = {2025-03-10},
  volume = {282},
  year = {1998}
}

@article{korshynska_generalized_2024,
  author = {Korshynska, Kateryna and Ulbricht, Sebastian},
  doi = {10.1103/PhysRevA.109.043321},
  journal = {Phys. Rev. A},
  month = {April},
  number = {4},
  pages = {043321},
  title = {Generalized {Josephson} effect in an asymmetric double-well potential at finite temperatures},
  url = {https://link.aps.org/doi/10.1103/PhysRevA.109.043321},
  urldate = {2025-03-10},
  volume = {109},
  year = {2024}
}

@article{PhysRevA.102.013325,
  author = {Pascucci, F. and Salasnich, L.},
  doi = {10.1103/PhysRevA.102.013325},
  issue = {1},
  journal = {Phys. Rev. A},
  month = {Jul},
  numpages = {7},
  pages = {013325},
  publisher = {American Physical Society},
  title = {Josephson effect with superfluid fermions in the two-dimensional BCS-BEC crossover},
  url = {https://link.aps.org/doi/10.1103/PhysRevA.102.013325},
  volume = {102},
  year = {2020}
}

@article{Vianello2024-ed,
  author = {Vianello, Cesare and Salasnich, Luca},
  copyright = {https://creativecommons.org/licenses/by/4.0},
  doi = {10.1038/s41598-024-65897-2},
  journal = {Sci. Rep.},
  month = {July},
  number = {1},
  pages = {15034},
  publisher = {Springer Science and Business Media LLC},
  title = {Condensate and superfluid fraction of homogeneous Bose gases in
a self-consistent Popov approximation},
  url = {https://doi.org/10.1038/s41598-024-65897-2},
  volume = {14},
  year = {2024}
}

@misc{bernhart2024,
  archiveprefix = {arXiv},
  author = {Bernhart, Erik and R{\"o}hrle, Marvin and Singh, Vijay Pal and Mathey, Ludwig and Amico, Luigi and Ott, Herwig},
  doi = {10.48550/arXiv.2409.03340},
  eprint = {2409.03340},
  month = {September},
  number = {arXiv:2409.03340},
  primaryclass = {cond-mat},
  publisher = {arXiv},
  title = {Observation of {{Shapiro}} Steps in an Ultracold Atomic {{Josephson}} Junction},
  urldate = {2025-09-29},
  year = {2024}
}

@article{pezze2024,
  author = {Pezz{\`e}, Luca and Xhani, Klejdja and Daix, Cyprien and Grani, Nicola and Donelli, Beatrice and Scazza, Francesco and {Hernandez-Rajkov}, Diego and Kwon, Woo Jin and Del Pace, Giulia and Roati, Giacomo},
  copyright = {2024 The Author(s)},
  doi = {10.1038/s41467-024-47759-7},
  issn = {2041-1723},
  journal = {Nature Communications},
  langid = {english},
  month = {June},
  number = {1},
  pages = {4831},
  publisher = {Nature Publishing Group},
  title = {Stabilizing Persistent Currents in an Atomtronic {{Josephson}} Junction Necklace},
  urldate = {2025-09-29},
  volume = {15},
  year = {2024}
}

@misc{singh2025,
  archiveprefix = {arXiv},
  author = {Singh, Vijay Pal and Bernhart, Erik and R{\"o}hrle, Marvin and Ott, Herwig and Mathey, Ludwig and Amico, Luigi},
  doi = {10.48550/arXiv.2509.03591},
  eprint = {2509.03591},
  month = {September},
  number = {arXiv:2509.03591},
  primaryclass = {cond-mat},
  publisher = {arXiv},
  title = {Weak-Link to Tunneling Regime in a {{3D}} Atomic {{Josephson}} Junction},
  urldate = {2025-09-29},
  year = {2025}
}

@book{zng,
  author = {Griffin, Allan and Nikuni, Tetsuro and Zaremba, Eugene},
  googlebooks = {yjDRQ\_e5UZIC},
  isbn = {978-1-139-47383-5},
  publisher = {Cambridge University Press},
  title = {Bose-{{Condensed Gases}} at {{Finite Temperatures}}},
  year = {2009}
}
\end{document}